\documentclass[twocolumn,aps,prd,amssymb,eqsecnum,nofootinbib]{revtex4}
\usepackage{amsmath}
\usepackage{amssymb}
\usepackage{epsfig}
\usepackage{bm}
\usepackage{slashbox}

\usepackage{color}
\usepackage{graphicx}
\usepackage{epsfig}
\usepackage{psfrag}
\usepackage{bbm}
\usepackage{rotating}
\usepackage{multirow}
\usepackage{amsmath}

\def\gg{\mathfrak g}

\begin{document}

\title{Momentum Flow in Black Hole Binaries: I. Post-Newtonian Analysis
of the Inspiral and Spin-Induced Bobbing}

\author{Drew Keppel, David A.\ Nichols, Yanbei Chen and Kip S. Thorne }
\affiliation{Theoretical Astrophysics, California Institute of Technology,
Pasadena, CA 91125.}
\date{23 February, 2009}

\begin{abstract}
A brief overview is presented of a new Caltech/Cornell research program
that is exploring the nonlinear dynamics of curved spacetime in binary black hole
collisions and mergers, and of an initial project in this program aimed at elucidating
the flow of linear momentum in black-hole binaries (BBHs).   The ``gauge-dependence''
(arbitrariness) in the localization of linear momentum in BBHs is discussed, along with
the hope that the qualitative behavior of linear momentum will be gauge-independent.
Harmonic coordinates are suggested as a possibly preferred foundation for fixing
the gauge associated with linear momentum.  For a BBH or other compact binary,
the Landau-Lifshitz formalism is used to define the momenta of the binary's individual
bodies in terms of integrals over the bodies' surfaces or interiors, and define the momentum
of the gravitational field (spacetime curvature) outside the bodies as a volume integral
over the field's momentum density.  These definitions will be used in subsequent papers that 
explore the internal nonlinear dynamics of BBHs via numerical relativity.  This formalism is then
used, in the 1.5PN approximation, to explore momentum flow between a binary's 
bodies and its gravitational field during the binary's orbital inspiral.  Special attention
is paid to momentum flow and conservation associated with synchronous spin-induced
bobbing of the black holes, in the so-called ``extreme-kick configuration'' (where two
identical black holes have their spins lying in their orbital plane and antialigned). 
\end{abstract}


\maketitle

\section{Introduction:  Motivation and Overview}
\label{sec:intro}

\subsection{Motivation}

Since the spectacular breakthrough by Pretorius \cite{Pretorius1} in spring 2005, numerical relativists have been successfully simulating the inspiral, merger and ringdown of binary
black holes (BBHs).  Much effort is now going into extracting physical and astrophysical information
from these simulations.

Almost all of this effort takes an \textit{``S-matrix'' viewpoint:}  For chosen initial conditions (the two
holes' initial masses, vectorial spins and orbital elements), what is the final emitted
gravitational waveform and what is the final hole's mass, vectorial spin, and kick velocity?

Equally interesting, it seems to us, are the things these simulations can teach us about the
\textit{nonlinear dynamics of curved spacetime}.  This paper is the first
in a new research program by the Caltech/Cornell relativity and 
numerical-relativity research groups, aimed at exploring nonlinear spacetime
dynamics in BBHs.  

\subsection{Momentum Flow in Black-Hole Binaries}

Several sets of analytical tools already exist
for exploring fully nonlinear space-time dynamics, for example
dynamical horizons~\cite{DynamicalHorizons} and 
quasi-local energy/momentum and angular momentum~\cite{Quasilocal}.  
One of our goals is to develop additional analytical and quasi-analytical
tools and use them to extract physical insights from numerical
simulations. 
Our initial focus in this direction is on \textit{the
distribution and flow of linear momentum in 
strongly nonlinearly curved spacetimes}---with linear momentum 
defined via psudotensors, which arise from viewing general relativity as a 
nonlinear field theory in a flat auxiliary spacetime~\footnote{Although
as Chen, Nester and Tung have shown, that various formulations of pseudotensors
could also be motivated from quasi-local points of view~\cite{CNT}.}
  This
paper is the first in a series that will deal with this subject.  

An instructive example is the \textit{extreme-kick configation} in which 
two identical, spinning black holes are initially
in a (quasi-)circular orbit, with oppositely directed spins lying in the 
orbital plane (Fig.\ \ref{fig:Binary}).  As Campanelli, Lousto, Zlochower
and Merritt \cite{CLZM1,CLZM2} (henceforth CLZM) discovered and
Gonzalez et.\ al.\ \cite{HGSBH} helped flesh out, 
of all initial configurations, this one has the largest kick speed for the 
final black hole,~\footnote{For binaries in non-circular orbits, larger 
kick velocities have been observed by Healy et al.~\cite{Healy}.}
and it also exhibits intriguing orbital motions:

\begin{figure}
\includegraphics[width=0.75\columnwidth]{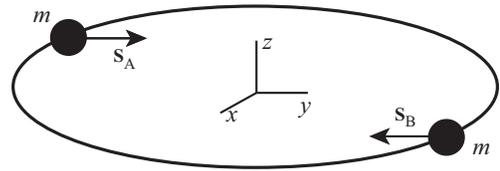}
\caption{
Extreme-kick configuration for a black-hole binary: Identical holes, $A$ and $B$ with masses
$m=M/2$ move in a circular orbit with their spin angular momenta $\bm S_A$ and
$\bm S_B$ antialigned and lying in the orbital plane.
}
\label{fig:Binary}
\end{figure}

\begin{figure}
\includegraphics[width=0.9\columnwidth]{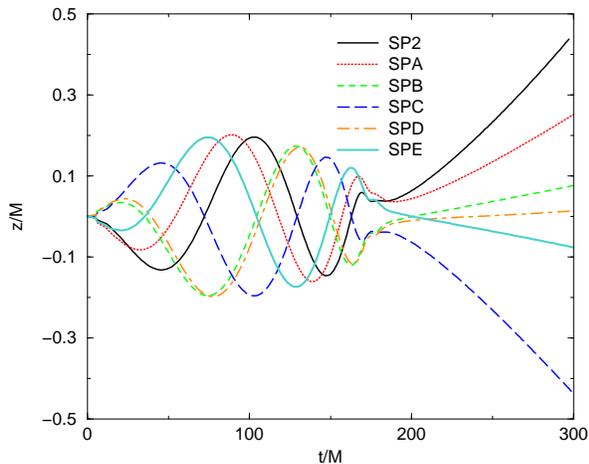}
\caption{
Bobbing and kick of binary black holes in the extreme-kick configuration of Fig.\ \ref{fig:Binary},
as simulated by 
Campanelli, Lousto, Zlochower and Merritt (CLZM) \cite{CLZM1}.  Plotted vertically (as a
function of time horizontally) is the identical 
height $z$ of the two black holes, and then transitioning through  merger (presumably at $t/M \sim 170$), the height of the merged hole, above the initial orbital plane.  This height versus time is shown for six different initial configurations, each leading to a different orbital phase
at merger.  In all six configurations, the initial holes' spins are half the maximum allowed, $a/m = 0.5$.
The height $z$ and time $t$ are those of the ``punctures'' that represent the holes'
centers in the CLZM computations, as defined in their computational coordinate system, which becomes
Lorentz at large radii. These $z$ and $t$ are measured in units of the system's 
total (ADM) mass $M \simeq
2m$.  
}
\label{fig:Bobbing}
\end{figure}

During the pre-merger inspiral, as the holes circle each other, they bob up and down (in the
$z$ direction of Fig.\ \ref{fig:Binary}), 
sinusoidally and synchronously.  After merger the combined hole gets kicked up or
down with a final speed that depends on the orbital phase at merger (relative to the spin directions). 
This bobbing then kick, as deduced by CLZM from numerical simulations, is graphed quantitatively
in Fig.\ \ref{fig:Bobbing}.  

Momentum conservation dictates that, when the holes are moving upward together with 
momentum $p_A^z + p_B^z$, there must be some equal and opposite downward 
momentum in their gravitational field (in the curved spacetime
surrounding them); and when the holes are moving downward, there must
be an equal and opposite upward field momentum.  How is this field momentum distributed,
and what are the details of the momentum flow between field and holes?  To what extent
is the final kick of the merged hole (e.g.\ in configuration SP2 of Fig.\ \ref{fig:Bobbing}) 
an inertial continuation of the holes' immediate pre-merger bobbing, and
correspondingly to what extent is the burst of 
downward gravitational-wave momentum that accompanies the kick caused by near-zone,
bobbing field momentum continuing ``inertially'' on downward after merger?  And to what
extent are other momentum-flow processes responsible for the motion shown in Fig.\ 
\ref{fig:Bobbing}?   These are the kinds of questions we would like to answer by  
an in-depth study of momentum flow in BBHs.

\subsection{Gauge-Dependence of Momentum Flow; Landau-Lifshitz Formalism}

The momentum distribution and flow in a relativistic binary are tricky concepts, because
momentum conservation arises from, and requires, translation invariance of spacetime.
Spacetime is translation invariant when flat, but not, in general, when
curved.  Two 
key exceptions are: (i) Spacetime is locally translation invariant in the vicinity of any
event, and this leads to the local law of 4-momentum conservation ${T^{\alpha\beta}}_{;\beta}
= 0$ (where $T^{\alpha\beta}$ is the total stress-energy tensor of all nongravitational
particles and fields).  (ii) Around any isolated system, e.g.\ a BBH, spacetime
can be idealized as asymptotically translation invariant, and this leads to the definition
and conservation law for the system's total momentum (e.g., the binary's final kick momentum
is equal and opposite to the momentum carried off by gravitational waves).  However, inside
the binary the curvature of spacetime prevents one from defining a globally conserved momentum
density and flux in any generally covariant way.

Nevertheless, we are quite hopeful that momentum flow can be developed into a powerful
tool for physical intuition into BBHs, and into the nonlinear 
dynamical behavior of curved spacetime that is generated by collisions of 
spinning black holes.  To do so, however,
will require living with the fact that the momentum distribution and momentum flux
inside a binary cannot be generally covariant, i.e.\ they must be, in some sense, 
\textit{gauge-dependent}.  

There, in fact, is a long and successful history of physicists' building up physical intuition
with the aid of gauge-dependent concepts; and it is that history that gives us hope.  For
example, in Maxwell's flat-spacetime electrodynamics, 
the vector potential satisfies the wave equation only if one first imposes Lorenz gauge;
and our physical intuition about electromagnetic waves relies, to a considerable extent,
on Lorenz-gauge considerations.  Similarly, in developing post-Newtonian ephemerides
for the solar system, celestial mechanicians have chosen a specific gauge in which to
work, and their intuition about relativistic effects in the solar system relies to a great extent
on that gauge's gauge-dependent constructs.  The choice of gauge was, to some extent,
arbitrary; but once the choice was made, intuition could start being built.  As a third
example, in black-hole perturbation theory relativists have built up physical intuition based
on Regge-Wheeler gauge, and based on the Teukolsky equation, each of which are 
gauge-dependent constructs.

The density, flux and conservation of linear momentum, in the curved spacetime of a black-hole
binary, must rely explicitly or implicitly on a mapping of the binary's curved spacetime
onto an auxiliary, translation-invariant flat spacetime.  This reliance is spelled out explicitly
in a \textit{reformulation of general relativity as a nonlinear field theory
in flat spacetime} presented in   
Landau and Lifshitz's \textit{Classical Theory of Fields} \cite{LL62}.  (See also Chap.\ 20 of
MTW \cite{MTW} and a more elegant, covariant formulation of the formalism developed by
Babak and Grishchuk \cite{Babak}.)  In the original Landau-Lifshitz formulation, one chooses any 
asymptotically Lorentz coordinates that one wishes, one maps onto an auxiliary flat spacetime
by asserting that these chosen (``preferred'') coordinates are globally Lorentz in the 
auxiliary spacetime (so in them
the auxiliary metric has components diag$[-1,1,1,1]$), 
and one then reformulates
the Einstein equations as a nonlinear field theory in the space of that flat, auxiliary metric.  The
result is a total stress-energy tensor 
\begin{equation}
\tau^{\alpha\beta} = (-g) (T^{\alpha\beta} + t^{\alpha\beta}_{\rm LL})\;,
\label{eq:tauDef}
\end{equation}
where $g$ is the determinant of the covariant components of the physical metric, $T^{\alpha\beta}$ is
the nongravitational stress-energy tensor, and $t^{\alpha\beta}_{\rm LL}$ is the ``Landau-Lifshitz
pseudotensor''.   By virtue of the translation invariance of the auxiliary spacetime, 
this $\tau^{\alpha\beta}$ has vanishing coordinate
divergence ${\tau^{\alpha\beta}}_{,\beta} = 0$ in the chosen ``preferred'' coordinates. Equivalently, 
this $\tau^{\alpha\beta}$ has vanishing 
covariant divergence ${\tau^{\alpha\beta}}_{|\beta} = 0$
with respect to the auxiliary flat metric.  The components $\tau^{j0}$ then represent the 
density and $\tau^{jk}$ the flux of a conserved linear momentum.

We envision each numerical relativity group choosing the coordinates used in its simulations to be
the ``preferred'' coordinates of this mapping to flat spacetime, resulting in each group's
adopting a different ``gauge''.  If we are lucky, this will
lead to momentum distributions and flows in different groups' simulations that are qualitatively
and semi-quantitatively 
similar.  If that is not the case, then we advocate that the community adopt, as a communally-agreed-upon
``preferred'' coordinate system (and thence gauge), Harmonic coordinates --- though even then
it might be necessary to face
up to the fact that Harmonic coordinates are not uniquely defined until one gives appropriate
initial conditions.  We envision joint numerical and quasi-analytical explorations, over the coming
months, that lead simultaneously to a choice or choices of ``preferred coordinates'' for the mapping
to flat spacetime, and physical insights into the flow of momentum in BBHs.

This paper represents a first small step in this direction:  To ensure that we understand quite
clearly what is going on, we shall focus in this paper on a binary's pre-merger bobbing, and we shall
study it and its momentum flow using the post-Newtonian approximation to general 
relativity in Harmonic coordinates.  Subsequent papers in this series will use the
Landau-Lifshitz formalism to explore momentum
flow in black-hole mergers.

\subsection{Overview of this Paper}
\label{sec:Overview}

We begin our post-Newtonian analysis in Sec.\ \ref{sec:ExtremeKick}
by presenting our main ideas and results in the simplest interesting context:
the extreme-kick configuration.  

We then, in the remainder of the paper, present a detailed post-Newtonian analysis of
spin-induced momentum flow in the inspiral phase of generic
compact binaries (BBHs, neutron-star binaries, or neutron-star / black-hole binaries).   
This detailed analysis begins in Sec.\ \ref{sec:LL} with a very brief summary of
the Landau-Lifshitz formalism, followed in Sec.\ \ref{sec:NonlinearBinary} by a use of the formalism
to give a general treatment of 4-momentum conservation for a fully 
relativistic system of compact bodies.  We express the binary's total 4-momentum
(as measured gravitationally by distant observers) as the sum of the 4-momenta
of its two bodies (expressed as integrals over their surfaces or, for stars,
volume integrals over their interiors) and the 4-momentum of their external gravitational field 
(expressed as a volume integral over the exterior).  We also derive expressions for the
rate of change of the 4-momentum of each body as a surface integral of the 
flux of 4-momentum being exchanged between the body and the external field.

In Sec.\ \ref{sec:NSBdetails} we specialize to the inspiral of a generic compact binary,
as analyzed in Harmonic coordinates
at leading nontrivial post-Newtonian order (1.5PN for the effects of spin); and we focus on the distribution and flow of linear momentum (the spatial part of 4-momentum)
induced by the bodies' spins.  
We begin in Sec.\ \ref{sec:ExtFieldMom} by computing the spin-induced perturbation of the field momentum
$\delta  \tau^{0j}$ in terms of the binary's masses, vectorial spins, and geometry; and
we then integrate this density over the exterior of the bodies to obtain the total field-momentum
perturbation $\delta \bm p_{\rm field}$ in terms of the bodies' masses, spins and vectorial
separation.  In Sec.\ \ref{sec:CMeom} we discuss the definition of a body's center of mass $\bm x_{\rm cm}$
and corresponding velocity $\bm v = d \bm x_{\rm cm}/dt$, and we
write down the influence of the bodies' spins $M d \delta \bm v/dt$ on their equations of motion.
In Sec.\ \ref{sec:BodyMomentum} and Appendix \ref{App:BHmom} we use our definition of center of mass to deduce an expression for the 
spin-induced perturbation of a body's momentum $\delta \bm p$ in terms of its mass times
velocity perturbation $M \delta \bm v$, and cross products of the bodies' spins with their 
separation vector.  Finally, in Sec.\  \ref{sec:momcons} we verify momentum conservation; i.e.,
we verify that, 
as the binary evolves and momentum is fed back and forth between the bodies and the field,
the bodies' equations of motion ensure that the spin-induced perturbation of the total momentum
(bodies plus field) is conserved.

\section{Bobbing and Momentum Flow in the Extreme-Kick Configuration}
\label{sec:ExtremeKick}

In this section we shall present an overview of our momentum-flow ideas and results
in the context of the extreme-kick configuration (Figs.\ \ref{fig:Binary} and \ref{fig:Bobbing}).

\begin{figure*}
\includegraphics[width=0.9\textwidth]{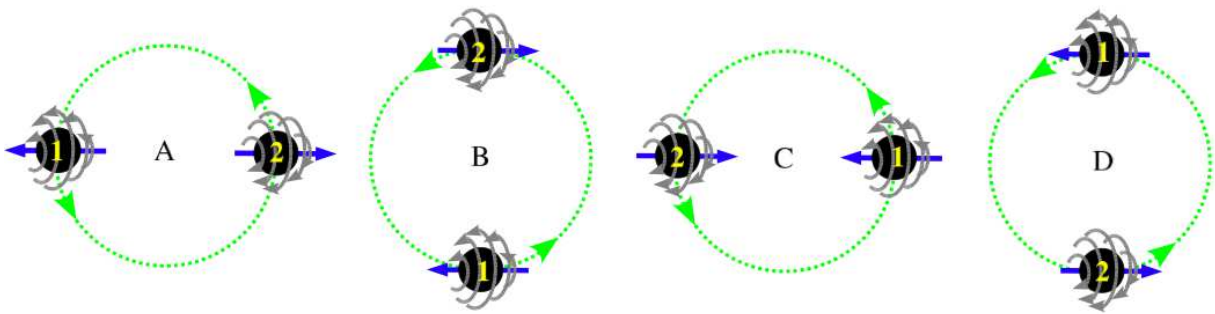}
\caption{Pretorius' physical explanation for the holes' bobbing
in the extreme-kick configuration.}
\label{fig:PretoriusIntuition}
\end{figure*}

Pretorius \cite{Pretorius2} has offered a lovely physical explanation for the holes' bobbing
(Fig.\ \ref{fig:Bobbing}) 
in this configuration:  In Fig.\ \ref{fig:PretoriusIntuition}, taken from his paper,
we see snapshots of the holes at four phases in their orbital motion.  In each snapshot,
each hole's spin drags space into motion (drags inertial frames) in the direction depicted by
gray, semi-circular arrows.  In phase B, hole 1 drags space and thence hole 2 into the
sheet of paper (or computer screen); and hole 2 drags space and thence hole 1 also 
inward.\footnote{This is very similar to the way that two fluid vortices (e.g.\ an aerofoil's 
starting and stopping vortex pair) drive each other into motion; see, e.g.\ \cite{Vortices}.}
In phase D each hole drags the other outward.  This picture agrees in phasing and 
semi-quantitatively in amplitude with the bobbing observed in the simulations (Fig.\ \ref{fig:Bobbing}).

Semi-quantitatively but not quantitatively.  In addition to frame dragging, there is a second
influence of the holes' spins on their motions, at the same 1.5 post-Newtonian (1.5PN)
order:  a force on each body due to the coupling of its own spin to the Riemann curvature
tensor produced by the other body.  For the extreme-kick configuration (Fig.\ 
\ref{fig:Binary}), in Harmonic coordinates the mass times coordinate acceleration produced by frame dragging (expressed as 
a weak perturbation, ``$\delta$'', to the motion of a nonspinning binary) is 
$( d^2  \delta \bm x_A / dt^2)_{\rm FD} = (4/r_{AB}^3) \bm S_B \times \bm v_{AB}$, and that produced
by spin-curvature coupling is $( d^2  \delta \bm x_A / dt^2)_{\rm SC} = (3/r_{AB}^3) \bm S_A \times \bm v_{AB}$
(first and second lines of Eq.\ (4.11c) of \cite{ThorneHartle}).  Here $r_{AB}$ is the separation between
the two holes and $\bm v_{AB} = \bm v_A - \bm v_B = 2 \bm v_A$ is the coordinate velocity of hole A relative to hole B.  
Since $\bm S_B = - \bm S_A$, the sum of the frame-dragging acceleration 
and spin-curvature-coupling acceleration is 
\begin{equation}
\left(\frac{d^2 \delta \bm x_A}{dt^2}\right)_{\textrm{spin effects}} = 
\frac{2}{r_{AB}^3} \bm S_A\times \bm v_A\;.
\label{fig:EKacceleration}
\end{equation}
We get the acceleration of hole B by replacing all subscript B's by subscript A's. The two holes' 
accelerations are identical (synchronous bobbing) because 
$\bm S_B= - \bm S_A$, $\bm v_B =-\bm v_A$.  

We can easily integrate this equation in time by noting that the spin precesses much
more slowly than the orbital motion so $\bm S_A$ can be approximated as constant, and noting that  
$\bm v_A$ rotates with angular velocity $\Omega = \sqrt{M/r_{AB}^3} = \sqrt{2m/r_{AB}^3}$, where $M$ 
is the total mass and $m$ is the mass of each hole.  The result, after one integration, can
be written as
\begin{equation}
m \delta \bm v_A = m \delta \bm v_B = -\frac{m}{r_{AB}^2} \bm S_A \times \bm n_{AB}\;.
\label{eqn:EKdeltav}
\end{equation}
Here $\delta \bm v_A$ and $\delta \bm v_B$ are the holes' velocity perturbations (bobbing)
produced by their spins, and  
$\bm n_{AB} = (\bm x_A - \bm x_B)/r_{AB}$ 
is the unit vector that points from hole B toward hole A. (As with
the convention for propagators, we regard things as ``moving'' from right to left; the
vector $n_{AB}$ points from B to A.)  
One might think that expression (\ref{eqn:EKdeltav}) represents the holes' bobbing {\it momentum}, 
but as we shall see in
Sec.\ \ref{sec:EKholemomenta} it does
not---for subtle but physically understandable reasons.  The bobbing momentum of each hole
is actually 2/3 times expression (\ref{eqn:EKdeltav}); see Eq.\ (\ref{eqn:deltapA}) below.

We can integrate Eq.\ (\ref{eqn:EKdeltav}) once more to obtain each hole's bobbing displacement
(change of location) in Harmonic coordinates.  We write the result in a form that is easily
compared with Fig.\ \ref{fig:Bobbing}:
\begin{equation}
\frac{\delta \bm x_A}{m} = - \bm v_A \times \frac{\bm S_A}{m^2}\;.
\label{eqn:EKDisplacement}
\end{equation}
Because $\bm S_A$ remains approximately constant while $\bm v_A$ rotates
uniformly in time (if we ignore radiation-reaction-induced inspiral), 
and because $\bm S_A$ and $\bm v_A$ both lie in the orbital plane,
Eq.\ (\ref{eqn:EKDisplacement}) represents an approximately sinusoidal bobbing orthogonal
to the orbital plane ($z$ direction), with (peak-to-peak) amplitude $\delta z /M = v_A S_A / m^2 = 
\frac12 v_A$, where we have used the spin magnitude $S_A/m^2 = a/m = 0.5$ of the
CLZM simulations.  

The CLZM simulations cover only the last two orbits before inspiral, when
the Post-Newtonian approximation is failing badly and the inspiral is rapid.  Nevertheless,
we can hope for rough quantitative agreement.  
The simulation shows a maximum bobbing amplitude
$\delta z/M \simeq 0.4$, which agrees  with our 1.5PN amplitude 
 $\delta z /M = \frac12 v_A$ if $v_A$ is near the speed of light, as it should be just before merger.
Half an orbit earlier the simulation's bobbing amplitude is smaller
by about a factor $1/1.7 \simeq 1/\sqrt{2.5}$, which is what our 1.5PN formula predicts
if the orbital radius is 2.5 times larger than at maximum amplitude - and this agrees 
fairly well with Fig.\ 2 of CLZM \cite{CLZM1}. 

\subsection{Field Momentum in the Extreme-Kick Configuration}
\label{sec:FieldMom}

In Harmonic gauge at leading post-Newtonian order, the Landau-Lifshitz formalism gives
for the density of field momentum
\begin{equation}
\tau^{0j} \bm e_j = -\frac{\bm g \times \bm H}{4\pi}
\label{eqn:gxH}
\end{equation}
(Eq.\ (4.1a) of \cite{KNT}).  Here, to the accuracy we need, $\bm g$ is the Newtonian gravitational
acceleration field (the \textit{gravitoelectric field}), $\bm H$ is the gravitational analog of the
magnetic field (the \textit{gravitomagnetic field}), and $\bm e_j$ is the $j$'th basis vector
of the flat-spacetime field theory that we are using.  (As we shall see, 
and as is discussed in Ref.\ \cite{KNT} and references cited therein, the analogy with electrodynamics
can be very powerful in building up insight into gravitational momentum density and flux.)  

[{\it Side Remarks:} 
Before applying Eq.\ (\ref{eqn:gxH}) to the extreme-kick configuration, let us build up 
a bit of physical insight into
it:  As for a particle, so for the relativistic gravitational field, we can regard
the ratio of the momentum density to the mass-energy density, $\tau^{0j}/\tau^{00}$ as a
field velocity 
\begin{equation}
v_{\rm field}^j \equiv \frac{\tau^{0j}}{\tau^{00}}\;.
\label{eqn:vfield}
\end{equation}
One can show that in Harmonic coordinates the vacuum field momentum density 
$\tau^{00}$ is negative; in fact, it is
\begin{equation}
\tau^{00} = - \frac{7}{8\pi}\bm g \cdot \bm g
\label{eqn:tau00}
\end{equation}
at leading PN order.\footnote{This is $\tau^{00}$ in Harmonic coordinates (gauge) in vacuum,
at leading (Newtonian) order; see, e.g.\ the first term in Eq.\ (4.4a) of \cite{PatiWill}.  This Newtonian
gravitational energy density is gauge dependent; see, e.g., the discussion in Box 12.3 of
\cite{BT}.}
Correspondingly, \textit{the gravitational field's velocity (as ``seen'' in our auxiliary flat
spacetime) points in the direction
of $+ \bm g \times \bm H$ and has a magnitude of order $|\bm H|/|\bm g|$.}  
The direction of this field velocity is the same as the direction of motion of an inertial point mass
(relative to our Harmonic coordinates) that is induced by a brief \textit{joint} action of $\bm g$ followed
by $\bm H$:  The geodesic equation, in Harmonic coordinates and for low particle velocities
$\bm v$ takes the Lorentz-force form
\begin{equation}
\frac{d{\bm v}}{dt} = \bm g + \bm v \times \bm H
\label{eqn:LorentzForce}
\end{equation}
(Eq.\ (2.7) of \cite{KNT}) at leading order.  
In a very short time interval $t$, the field $\bm g$ acting on a particle initially at rest
produces a velocity $\bm v = \bm g t$, and then
$\bm H$ acts on this velocity to produce $\delta \bm v = \frac12 \bm g \times \bm H t^2$ --- which
points in the direction of the field velocity.]

Now let us study the field momentum for the extreme-kick black-hole binary.  
For the moment we are only interested in that portion of the field momentum which is induced
by the holes' spins, since this is the portion that must flow back and forth between the field and
the bobbing holes in order to conserve total momentum.  This portion arises from one hole's  gravitoelectric field $\bm g$ coupling to the spin-induced part of the the other hole's gravitomagnetic field
\begin{equation}
\delta \tau^{0j} \bm e_j = 
- \frac{\bm g_A \times \bm H^{\rm spin}_B}{4\pi} 
- \frac{\bm g_B \times \bm H^{\rm spin}_A}{4\pi} \;.
\label{eqn:deltatau0j}
\end{equation}
Here
\begin{subequations}
\begin{eqnarray}
\bm g_A &=& - \frac{m}{r_A^2} \bm n_A\;,
\label{eqn:gA} \\
\bm H_A &=& - 2 \frac{(3 \bm n_A \cdot \bm S_A) \bm n_A - \bm S_A}{r_A^3}
\label{eqn:HAspin}
\end{eqnarray}
\label{eqn:gHspin}
\end{subequations}
(Eqs.\ (2.5) and (6.1) of \cite{KNT}), with $\bm n_A$ the unit radial vector pointing from
the center of hole $A$ to the field point, and $r_A$ the distance from the center of hole $A$ to the field point.  The gravitoelectric field (\ref{eqn:gA}) (actually the Newtonian gravitational acceleration) has
identically the same form as the Coulomb electric field of a point charge, with the charge replaced
by the hole's mass $m_A$ and the sign reversed.  Similarly, the gravitomagnetic field
(\ref{eqn:HAspin}) is identical to the dipolar magnetic field of a point magnetic dipole, with the
magnetic moment replaced by the opposite of twice the hole's spin, i.e., $-2 \bm S_A$.  The fields
for hole $B$ are the same as Eqs.\ (\ref{eqn:gHspin}), but with each subscript $A$ replaced by
a $B$.  

Combining Eqs.\ (\ref{eqn:deltatau0j}) and (\ref{eqn:gHspin}) we obtain for the binary's density of
field momentum (that portion which must flow during bobbing 
\footnote{There are portions of the momentum density that do not flow during bobbing, which will be
important for our comparisons with numerical relativity.
The full expression, therefore, is listed in Appendix \ref{sec:MomDenFull}.})
\begin{eqnarray}
\delta\tau^{0j}{\bm e}_j &=& \frac{m}{2\pi r_A^3 r_B^2} \left[ 3({\bf S}_A \cdot {\bf n}_A)({\bf n}_A \times {\bf n}_B) - 
({\bf S}_A \times {\bf n}_B)\right]\nonumber\\
&& + (A\leftrightarrow B),
\label{eqn:EKfieldmom}
\end{eqnarray}
and integrating this over the space outside the holes, we obtain for the total field momentum that
flows during bobbing
(the part of the field momentum that depends on the holes' spins)\footnote{Equations (\ref{eqn:EKfieldmom})
and (\ref{eqn:EKdeltapfield}) are special cases of Eqs.\ (\ref{eqn:tau0jderivs}) and (\ref{pjext_final}) 
below, where the details of the integration are
carried out.}
\begin{equation}
\delta \bm p_{\rm field} = \frac43 \frac{m}{r_{AB}^2} \bm S_A \times \bm n_{AB}\;.
\label{eqn:EKdeltapfield}
\end{equation}

This is equal and opposite to the sum of the holes' bobbing momenta $\delta p_A + \delta p_B$,
as we shall see in Sec.\ \ref{sec:EKMomCons} below.

\begin{figure}
\includegraphics[width=1.0\columnwidth]{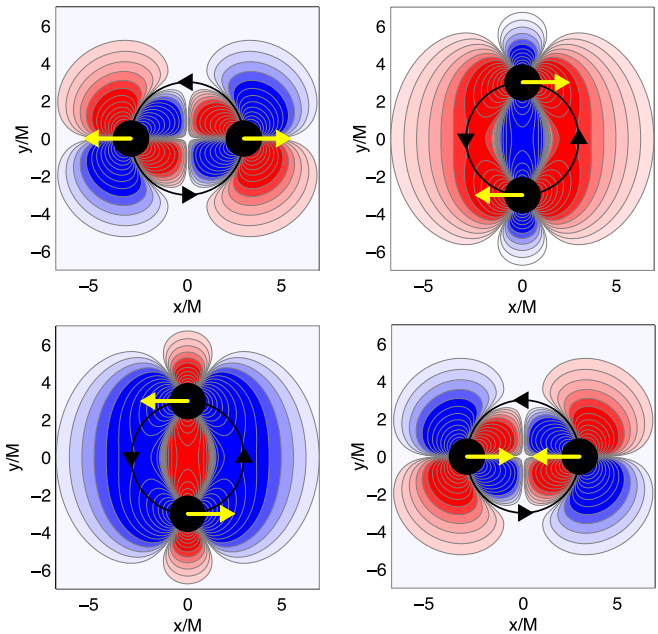}
\caption{The four pictures show the $z$-component of field-momentum density $\delta \tau^{0z}$ in the orbital plane at
four different times, a quarter orbit apart.  
Red represents positive momentum density (coming out of the paper), and blue, negative (going into
the paper).
Only the piece of momentum density  $\delta \tau^{0z}$ that flows during bobbing [Eq.\ (\ref{eqn:EKfieldmom})]
is depicted.  
The yellow arrows are the black holes' vectorial spins; the large, black arrowed circle shows 
the orbital path of the two holes.
In the top-left picture, one sees the density of momentum when the black holes are at the top of their
bob (maximum $z$) and momentarily stationary [Eqs.\ (\ref{eqn:EKdeltav}) and 
(\ref{eqn:EKDisplacement})]. 
The gravitational-field momentum is zero, but the momentum density itself shows rich structure.
A quarter orbit later, in the top right, the holes are moving downward (into the paper) at top
speed. 
The momentum between the black holes (blue region) flows into the paper with them, while 
surrounding momentum (red region) flows out of the paper ($+z$ direction).
A half orbit after the first picture, in the lower left, 
the holes are momentarily at rest at the bottom of their 
bob (minimum $z$), the net field momentum is zero, and the momentum distribution is opposite that
in the first picture  (as one would expect during sinusoidal bobbing).
Similarly, three quarters of the way through the orbit, in the lower right, the holes have reached 
their maximum upward speed, and the momentum distribution is identical to the
second figure, but with the opposite sign.}
\label{fig:MomDensity}
\end{figure}

Figure \ref{fig:MomDensity} shows the $z$-component (perpendicular to the orbital plane) of the field-momentum 
density $\delta \tau^{0z}$, as measured in the orbital plane at four different moments in the binary's orbital evolution.
Only that part of the momentum which flows during bobbing [Eq.\ (\ref{eqn:EKfieldmom})] is pictured.
Red depicts momentum density flowing out of the paper ($+z$ direction), and blue, into the paper.
The yellow arrows show the holes' vectorial spins $\bm S$, and the arrowed circle is the binary's orbital trajectory.
In the top-left and bottom-right frames, the black holes are momentarily stationary at the top and bottom of their bobbing [cf.\ 
Eqs.\ (\ref{eqn:EKdeltav}) and
(\ref{eqn:EKDisplacement})].  
Nevertheless, the momentum density has a non-trivial distribution.
In the top-right and bottom-left frames, the black holes are moving downward and upward, respectively, with maximum speed.
In both cases, the field-momentum density between the two holes flows in the same direction as the bobbing,
whereas the momentum surrounding the binary is in the opposite direction and larger.
This leads to net momentum conservation for the binary, as discussed in Section \ref{sec:momcons}.

It is worth noting that the four figures, going counterclockwise from the top-left, are taken a quarter period 
apart in orbital phase.
The first and third differ by half an orbital period (as do the second and fourth); and, consequently, 
the momentum patterns of each pair are identical, but signs are reversed (red exchanged with blue, as dictated by
the symmetry of the configuration). 
This feature is responsible for the sinusoidal bobbing.

\subsection{The holes' momenta}
\label{sec:EKholemomenta}

In this section we shall use a roundabout route to explain, physically, why the momentum
$\bm p_A$ of black hole $A$ is not $m \bm v_A$, and to derive an expression for it.

Begin by considering, for pedagogical purposes, a rigidly and slowly rotating body in
flat spacetime with rotational velocity $\bm v_{\rm rot}(\bm x)$.  
Let $\rho$ be the mass-energy density of the body's material in the local
rest frame of a bit of material.  Then \textit{in an inertial frame where the body is at rest except
for its rotation} (the body's ``momentary inertial frame''), its mass-energy density is 
$T^{00} = \rho (1+ v_{\rm rot}^2)$, where $\frac12 v_{\rm rot}^2$ comes from kinetic
energy and $\frac12 v_{\rm rot}^2$ from Lorentz contraction.
We define
the body's center-of-mass location $\bm x_{\rm cm}$ by  
\begin{equation}
M \bm x_{\rm cm} = 
\int T^{00} \bm x d^3x \quad \textrm{in body's momentary rest frame}\;,
\label{eq:xcmDef}
\end{equation}
where $M \equiv \int T^{00} d^3x$ is the body's mass.  If the body is
weakly gravitating, this location will be the center of the monopolar part of its gravitational
field.  

Now let this rotating body move with a linear velocity $\bm v$ that is small compared
to its rotational velocity, so $T^{00} = \rho [1
+ (\bm v + \bm v_{\rm rot})^2] \simeq \rho[1+\bm v_{\rm rot}^2 + 2 \bm v_{\rm rot} \cdot \bm v]$.
If we use this $T^{00}$ to compute $\int T^{00} \bm x d^3x$, we will not get the $\bm x_{\rm cm}$ 
of Eq.\ (\ref{eq:xcmDef}) because
the term $2\rho \bm v_{\rm rot} \cdot \bm v$ will weight $\bm x$ extra heavily on the side
of the body where $\bm v_{\rm rot} \cdot \bm v >0$ and less heavily on the side where
$\bm v_{\rm rot} \cdot \bm v <0$.  If we want to compute the correct $\bm x_{\rm cm}$ by an
integral performed in a frame where the body is moving, we must correct for this effect.  The correction
factor is well known (\cite{SSC} and Sec.\ \ref{sec:CMeom} below):
\begin{equation}
M \bm x_{\rm cm}  = \int T^{00} \bm x d^3 x - \bm v \times \bm S\;,
\label{eqn:xcm}
\end{equation}
where $\bm S$ is the body's angular momentum.   Other definitions of center-of-mass
are sometimes used, but they all differ from the locations one would identify, in the body's
rest frame, as the mass-energy-weighted location (\ref{eq:xcmDef}) and the 
center of weak monopolar
gravity---i.e., they are less \textit{physically motivated}
than this one.  

Equation (\ref{eqn:xcm}), when extended into general relativity in the obvious
manner,
\begin{equation}
M \bm x_{\rm cm}  = \int \tau^{00} \bm x d^3 x - \bm v \times \bm S\;
\label{eqn:xcmGR}
\end{equation}
[with $\tau^{\alpha\beta}$ the total stress-energy tensor of Eq.\ (\ref{eq:tauDef})], is called the \textit{physical
spin supplementary condition}~\cite{SSC}; cf.\ Eq.\ (\ref{eqn:xcmp0}) below, where a formal 
derivation is presented.    
In general relativity this condition guarantees that
in the body's local rest frame, 
$\bm x_{\rm cm}$ is at the center of the monopolar part of the body's (possibly strong) gravitational
field, or more precisely the center of the monopolar part of the time-time component of the metric density
$\mathfrak{g}^{00} \equiv \sqrt{-g} g^{00}$, which plays a major role in the Landau-Lifshitz
field-theory-in-flat-spacetime formalism.

The black-hole velocities $\bm v_A$ and $\bm v_B$ used in this paper and in the standard
Harmonic-coordinate, post-Newtonian equations of motion, are the (coordinate) time derivatives
of the holes' centers of mass: 
\begin{equation}
\bm v_A = d \bm x_{A\;\rm cm} /dt\;.
\label{eq:vADef}
\end{equation}

By specializing Eq.\ (\ref{eqn:xcmGR}) to body $A$, differentiating with respect to time,  using the conservation law 
${\tau^{00}}_{,0}+ {\tau^{0j}}_{,j} =0$,
and integrating by parts on the volume integral, we obtain [Eq.\ (\ref{eqn:pjvint}) below and its derivation]
\begin{eqnarray}
p^j_A &=& \underbrace{m v^j_A}_{\mbox{\small kinetic}} +  
\underbrace{(\bm a_A \times \bm S_A)^j}_{{\mbox{\small SSC}}} \nonumber \\  
&&+ \underbrace{\int_{\partial A}(x^j - x^j_{{\rm cm}\, A}) (\tau^{0k} - \tau^{00} v^k_A) d\Sigma_k}_{_{\mbox{\small surface}}} 
\;,
\label{eqn:pAj}
\end{eqnarray}
where 
\begin{equation}p^j_A \equiv \int_A \tau^{0j} d^3x
\label{eqn:pjAdefin}
\end{equation}
is the total 4-momentum of body $A$. 
Here  ${\bf a}_A$ is the acceleration of body $A$ produced
by the gravity of body $B$ and $m$ is the mass of body $A$.  
For a black hole the linear momentum must be defined via a surface integral rather than 
the volume integral $\int_A \tau^{0j} d^3x$ [Eq.\ (\ref{eqn:pAsurf}) below], but Eq.\ (\ref{eqn:pAj}) 
still turns out to be true; see the paragraph following Eq.\ (\ref{eqn:pjvint}) below. 

Equation (\ref{eqn:pAj}) has a physical interpretation that is closely related to the one
for the center-of-mass equation (\ref{eqn:xcmGR}) that underlies it:  Rearranged, this equation
says 
$m{\bf v}^j_A = p_A^j - ({\bf a}_A \times  {\bf S}_A)^j - \int_{\partial A}(x^j - x^j_{{\rm cm}\, A}) (\tau^{0k} - \tau^{00} v^k_A) d\Sigma_k$.
The left side is the time derivative of the center-of-mass location, weighted by the body's mass (or the kinetic momentum).  The
first term on the right side is the body's total momentum, 
 i.e. the volume integral of $\tau^{0j}$.  
 The second (SSC) term corrects for the fact that
for a spinning body $\tau^{0j}$ weights the center of mass too heavily on the side of the 
body where the rotational velocity and linear velocity are coaligned ($\bm v_{{\rm A}\;{\rm rot}} 
\cdot \bm v_{\rm A} >0$) and too lightly on the side where
they are antialigned.  The third (surface) term corrects for a contribution to the momentum arising from 
mass flowing into and out of the body (mass flux $\tau^{0i}-\tau^{00}v^i$) at different locations
on the body's face.

\subsection{Momentum Conservation for Extreme-Kick Configuration}
\label{sec:EKMomCons}

We now specialize Eq.\ (\ref{eqn:pAj}) for the momentum of a body in a generic binary to the extreme-kick
configuration and focus on the spin-dependent piece of the momentum that is exchanged between the bodies and the field.  
By inserting the expressions $\delta\tau^{0j} \bm e_j = - \bm g_B \times \bm H_A/4\pi$ and $\bm a_A
= \bm g_B(\bm x = \bm x_{A\;{\rm cm}})$ into
Eq.\ (\ref{eqn:pAj}), with $\bm g$ and $\bm H$ given by Eqs.\ (\ref{eqn:gHspin}), and performing the integrals, we obtain 
\begin{equation}
\delta \bm p_A = 
-\frac23 \frac{m}{r_{AB}^2} \bm S_A \times \bm n_{AB}\;.
\label{eqn:deltapA}
\end{equation}
This spin-induced perturbation of the linear momentum of body $A$, when added
to an equal amount for body $B$ gives $-\frac43 (m/r_{AB}^2) \bm S_A \times \bm n_{AB}$,
which is equal and opposite to the spin-induced perturbation of the field momentum
(\ref{eqn:EKdeltapfield}).  Therefore, as the holes circle each other, momentum flows
sinusoidally back and forth between the holes and the field, with no change in the
total momentum; the total momentum is conserved during the bobbing.

\renewcommand{\arraystretch}{2.5}
\begin{table*}
\begin{tabular}{|c|c|c|c|c|c|}
\hline
& \multicolumn{2}{|c|}{Kinetic} & &  &\\
\cline{2-3}
Body & Frame-Dragging & Spin-Curvature 
 & SSC & Surface & Total  \\
\hline \hline
$\bm p_A$ & $-4 \mathbf{a}_B \times \mathbf{S}_B$ & $3\mathbf{a}_A \times \mathbf{S}_A$ 
 & $\mathbf{a}_A \times \mathbf{S}_A$ &  $\displaystyle -\frac{2}{3}\mathbf{a}_A \times \mathbf{S}_A$ & $\displaystyle -\frac{2}{3}\mathbf{a}_A \times \mathbf{S}_A$ \\ 
\hline
$\bm p_B$ & $-4 \mathbf{a}_A \times \mathbf{S}_A$ & $3\mathbf{a}_B \times \mathbf{S}_B$
 & $\mathbf{a}_B \times \mathbf{S}_B$ &  $\displaystyle -\frac{2}{3}\mathbf{a}_B \times \mathbf{S}_B$ & $\displaystyle -\frac{2}{3}\mathbf{a}_B \times \mathbf{S}_B$ \\ 
\hline 
$\bm p_{\rm field}$ & &  & & &  $\displaystyle \frac{2}{3}(\mathbf{a}_A \times \mathbf{S}_A+\mathbf{a}_B \times \mathbf{S}_B)$\\
\hline
\end{tabular}
\caption{\label{tab} Spin-dependent, time-varying pieces of body and field momenta at 1.5PN order, for the extreme-kick binary 
(circular orbit with spins antialigned and in the orbital plane). The body momenta are broken down into kinetic, SSC and surface terms
and are expressed in terms of the bodies' spins $\bm S_{A,B}$ and Newtonian-order gravitational accelerations $\mathbf{a}_{A,B} = 
-m \bm n_{A,B}/r_{AB}^2$. 
See Eqs.~\eqref{eqn:eomTH} and~\eqref{deltapA3} for a similar decomposition in a generic binary.
}
\end{table*}


Let us examine in detail how momentum conservation is achieved in the presence of
the bodies' bobbing.  
Our detailed analysis (above) breaks each object's momentum perturbation $\delta \bm p_{A,B}$ into three terms, the 
{\it kinetic momentum} $m \delta \bm v_{A,B}$, a term due to the {\it SSC}, and a {\it surface} integral term (see Table~\ref{tab}). 
The total kinetic momentum 
\begin{equation}
m \delta\mathbf{v}_A + m \delta \mathbf{v}_B = - (\mathbf{a}_A \times \mathbf{S}_A + 
\mathbf{a}_B\times \mathbf{S}_B)
 \neq 0
 \end{equation} 
is not conserved because of the non-cancellation 
between the frame-dragging and spin-curvature coupling  
terms.  The total body momentum $\delta \bm p_A + \delta \bm p_B$ is not conserved either; it sums up to $2/3$ the total kinetic momentum:
\begin{equation}
\delta \mathbf{p}_A  +\delta \mathbf{p}_B =  -\frac{2}{3}(\mathbf{a}_A \times \mathbf{S}_A + 
\mathbf{a}_B\times \mathbf{S}_B)
 \neq 0,
 \end{equation} 
To achieve momentum conservation, there is a non-zero spin-dependent total field momentum distributed outside of the bodies, 
with
\begin{equation}
\delta \mathbf{p}_{\rm field} =  \frac{2}{3}(\mathbf{a}_A \times \mathbf{S}_A + 
\mathbf{a}_B\times \mathbf{S}_B)\,.
\end{equation}
Note that this total external field momentum is only $-2/3$ the spin-dependent total kinetic momentum --- instead of the $-1$ that one might have naively expected.

It is important to notice that for each body, a {\it canonical momentum} can be formed by adding the SSC term to the kinetic momentum
\begin{equation}
\bm p_{A\;{\rm canonical}} = m \bm v_A + \bm a_A\times  \bm S_A\;.
\label{eqn:Pcanonical}
\end{equation}
The total canonical momentum $\bm p_A + \bm p_B$ is conserved, because the sum of the surface terms of the bodies' momenta and the 
external field momentum are
equal and opposite.  (This is so not solely for our extreme-kick configuration, but also for any generic binary; see Sec.\ 
\ref{sec:NSBdetails} 
below).   This canonical momentum can be motivated quite simply by special relativistic kinetics, 
{\it without the need for any knowledge of field momentum,}
and it is used in the Hamiltonian approach to post-Newtonian dynamics~\cite{DJS,SSC}. 

Although the introduction of canonical momentum resolves the issue of momentum conservation at the 
level of two-body dynamics, it does not provide information about the distribution of field momentum nor 
the role of field momentum in momentum conservation.  Our analysis reveals substantial 
spin-dependent  field-momentum outside of the bodies --- with the same order of magnitude as the total spin-dependent kinetic mometum. 

One might question the meaningfulness of a distiction between the bodies' (localized) momenta and 
the (distributed) field momenta, because a different choice of gauge might move momentum between these two components, and 
possibly even move all the field momentum into the interiors of the objects.  
We  argue that our choice of Harmonic gauge, the analogue of Lorenz gauge in electrodynamics, is a promising tool for analyzing
compact binaries, since its metric perturbations (both physical and gauge) 
propagate at the speed of light, which will make gauge effects behave causally just as do physical effects.  For this reason, and because of the physical intuition that the above analysis brings, we
advocate using Harmonic gauge and its nonzero field momentum in analyzing compact, inspiraling binaries.


\section{The Landau-Lifshitz Formalism in Brief}
\label{sec:LL}

We turn, now, to a detailed analysis of momentum flow in generic compact binary systems.
We begin in this section with a very brief review of the Landau-Lifshitz (LL) formulation of 
general relativity as a nonlinear field theory in flat spacetime \cite{LL62}.

This formulation starts (as discussed in Sec.\ \ref{sec:ExtremeKick} above) 
by introducing an (arbitrary) coordinate system in which
the auxiliary flat metric takes the Minkowski form $\eta_{\mu\nu} = {\rm diag}(-1,1,1,1)$.  
Gravity is described, in this formulation, 
by the physical metric density
\begin{equation}
\gg^{\mu \nu} = \sqrt{-g} g^{\mu \nu}\;.
\label{eqn:gg}
\end{equation}
Here $g$ is the determinant of the covariant components of the physical metric, and $g^{\mu\nu}$
are the contravariant components of the physical metric.  In terms of the superpotential
\begin{equation}
H^{\mu\alpha\nu\beta} \equiv \gg^{\mu \nu}\gg^{\alpha \beta} - \gg^{\mu \alpha}\gg^{\nu \beta}\;,
\label{eqn:superpotential}
\end{equation}
the Einstein field equations take the field-theory-in-flat-spacetime form
\begin{equation}
{H^{\mu\alpha\nu\beta}}_{,\alpha\beta} = 16\pi \tau^{\mu\nu}\;.
\label{eqn:efe}
\end{equation}
Here $\tau^{\mu\nu} = (-g)(T^{\mu\nu} + t^{\mu\nu}_{\rm LL})$ is 
the total effective stress-energy tensor introduced in Eq.\ (\ref{eq:tauDef}), indices
after the comma denote partial derivatives (covariant derivatives with respect to the flat
auxiliary metric), and the Landau-Lifshitz pseudotensor $t^{\mu\nu}_{\rm LL}$
(actually a real tensor in the
auxiliary flat spacetime) is given by Eq.\ (100.7) of LL \cite{LL62} or equivalently Eq.\ (20.22) of MTW
\cite{MTW}.  By virtue of the symmetries of the superpotential (which are the same as those
of the Riemann tensor), the field equations in the form (\ref{eqn:efe}) imply the differential
conservation law for 4-momentum
\begin{equation}
{\tau^{\mu\nu}}_{,\nu} = 0\;,
\label{eqn:divtau}
\end{equation}
which is equivalent to ${T^{\mu\nu}}_{;\nu} = 0$ (where the semicolon denotes a covariant 
derivative with respect to the physical metric).  

It is shown in LL and in MTW that the total 4-momentum of any isolated system (as measured 
gravitationally in the asymptotically flat region far from the system) is 
\begin{equation}
p^\mu_{\rm tot} = \frac{1}{16\pi}\oint_{\mathcal S} {H^{\mu\alpha 0 j}}_{,\alpha} d\Sigma_j\;,
\label{eqn:MomTot}
\end{equation}
where $d\Sigma_j$ is the surface-area element (defined, of course, using the flat auxiliary metric), and 
the integral is over an arbitrarily large closed surface $\mathcal S$ surrounding the system.  This
total 4-momentum satisfies the standard conservation law
\begin{equation}
\frac{dp^\mu_{\rm tot}}{dt} = - \oint_{\mathcal S} \tau^{\mu j} d\Sigma_j\;.
\label{eqn:dMomTotdt}
\end{equation}
[The proof of this given in LL and MTW relies on an assumption that the interior of $\mathcal S$ be
simply connected, i.e.\ that it not contain any black holes.  However, that assumption is not 
necessary:  Differentiate Eq. (\ref{eqn:MomTot}) with respect to $t$, then use 
${H^{\mu\alpha 0 j}}_{,\alpha 0} 
= {H^{\mu\alpha \nu j}}_{,\alpha \nu} - {H^{\mu\alpha k j}}_{,\alpha k}\;.$  The first term is
$-16\pi \tau^{\mu j}$ by virtue of the the field equations (\ref{eqn:efe}) and the antisymmetry
of the superpotential on its last two indices; and its surface integral gives the right side
of Eq.\ (\ref{eqn:dMomTotdt}).  That same antisymmetry on the second term
$- {H^{\mu\alpha k j}}_{,\alpha k}$ 
permits us to write it as the curl of a 3-vector field, whose surface integral vanishes by virtue
of Stokes' theorem.  The result is Eq.\ (\ref{eqn:dMomTotdt}).]

\section{4-Momentum Conservation for a Fully Nonlinear Compact Binary}
\label{sec:NonlinearBinary}

\begin{figure}
\includegraphics[width=0.6\columnwidth]{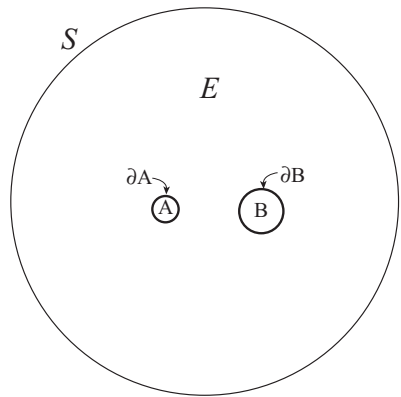}
\caption{The regions of space around and inside a compact binary system.}
\label{fig:Binary1}
\end{figure}

We now apply this LL formalism to a binary system made of black holes and/or neutron stars;
see Fig.\ \ref{fig:Binary1}.
We denote the binary's two bodies by the letters $A$ and $B$, and the regions of space inside them by
these same letters, and their surfaces by $\partial A$ and $\partial B$.  For a black hole,
$\partial A$ could be the hole's absolute event horizon or its apparent horizon, whichever one wishes. 
For a neutron star, $\partial A$ will be the star's physical surface.   We denote by 
$\mathcal E$ the region outside both bodies, but inside the arbitrarily large surface 
$\mathcal S$ where the system's total momentum is computed.  (In later papers in this
series, $\mathcal S$ will sometimes be the outer boundary of a numerical-relativity computational
grid.)  

By applying Gauss's theorem to Eq.\ (\ref{eqn:MomTot}) for the binary's total 4-momentum
and using the Einstein field equation (\ref{eqn:efe}), we obtain an expression
for the binary's total 4-momentum as a sum over contributions from each of the bodies
and from the gravitational field in the region $\mathcal E$ outside them: 
\begin{subequations}
\begin{equation}
p_{\rm tot}^\mu  = p_A^\mu + p_B^\mu + p_{\rm field}^\mu\;.
\label{eqn:pDecompose}
\end{equation}
Here
\begin{equation}
p_A^\mu \equiv  \frac{1}{16\pi}\oint_{\partial A} {H^{\mu\alpha 0 j}}_{,\alpha} d\Sigma_j\;
\label{eqn:pAsurf}
\end{equation}
is the 4-momentum of body A and similarly for body B, and 
\begin{equation}
p_{\rm field}^\mu \equiv \int_{\mathcal E} \tau^{0\mu} d^3 x\;.
\label{eqn:pEvol}
\end{equation}
is the gravitational field's 4-momentum in the surrounding space.

If either of the bodies has a simply connected interior (is a star rather than a black hole),
then we can use Gauss's theorem and the Einstein field equations (\ref{eqn:efe}) to convert the surface
integral (\ref{eqn:pAsurf}) for the body's 4-momentum into a volume integral over the body's interior:
\begin{equation}
p_A^\mu =   \int_{A} \tau^{0\mu} d^3 x\;.
\label{eqn:pAvol}
\end{equation}
\label{eqn:pComponents}
\end{subequations}

By an obvious extension of the argument we used to derive Eq.\ (\ref{eqn:dMomTotdt}) for the rate
of change of the binary's total 4-momentum, we can deduce from Eq.\ (\ref{eqn:pAsurf})
the corresponding equation for the rate of change of the 4-momentum of body A:
\begin{equation}
\frac{dp^\mu_A}{dt} = - \oint_{\partial A} (\tau^{\mu k} - \tau^{\mu 0} v^k_A ) d\Sigma_k\;.
\label{eqn:dpAdtsurf}
\end{equation}
Here the second term arises from the motion of the boundary of body $A$ with
coordinate velocity $v^k_A = dx_{A\;{\rm cm}}^k/dt$.  Equation (\ref{eqn:dpAdtsurf}) 
describes the flow of field 4-momentum into and out of body A.

We shall use Eqs.\ (\ref{eqn:pComponents}), (\ref{eqn:dMomTotdt}), and
(\ref{eqn:dpAdtsurf}), specialized to linear momentum (index $\mu$ made spatial)
as foundations for our study of momentum flow in compact binaries.  

The actual values of the body and field 4-momenta, computed in the above ways,
 will depend on the arbitrary coordinate
system that we chose, in which to make the auxiliary metric be diag[-1,1,1,1] and in which
to perform the above computations.  This is the ``gauge-dependence'' discussed above.  
In the remainder of this paper we shall choose Harmonic coordinates, so the 
gravitational field satisfies the Harmonic gauge condition
\begin{equation}
{\gg^{\alpha\beta}}_{,\beta} = 0\;,
\label{eqn:HarmonicGauge}
\end{equation}
and we shall specialize the above equations
to the 1.5 post-Newtonian
approximation and use them to study momentum flow during the inspiral phase of generic compact
binaries. In future papers we shall use the above equations, combined with numerical-relativity
simulations, to study momentum flow during the collision, merger, and ringdown
phases of compact binaries.

\section{Post-Newtonian Momentum Flow in Generic Compact Binaries}
\label{sec:NSBdetails}

\subsection{Field Momentum Outside the Bodies}
\label{sec:ExtFieldMom}

As for the extreme-kick configuration, so also in general, the portion of the field momentum that is induced by the bodies' spins and that flows back and forth between the field and the bodies, as the bodies move, is 
\begin{equation}
\delta \tau^{0j} \bm e_j = 
- \frac{\bm g_A \times \bm H^{\rm spin}_B}{4\pi} 
- \frac{\bm g_B \times \bm H^{\rm spin}_A}{4\pi} \;
\label{eqn:deltatau0jagain}
\end{equation}
[Eq.\ (\ref{eqn:deltatau0j})].  We find it convenient to rewrite the bodies' gravitoelectric and gravitomagnetic fields (\ref{eqn:gHspin}) as
\begin{equation}
g_K^j = m_K \left(\frac{1}{r_K}\right)_{,j}\;,\quad
H_K^j = -2 S_K^i \left(\frac{1}{r_K}\right)_{,ij}\;,
\label{eqn:gHderivs}
\end{equation}
where $K$ is $A$ or $B$ and where, as before, $r_K$ is the (flat-space) distance
of the field point from the center of mass of body $K$.  Inserting Eqs.\ (\ref{eqn:gHderivs}) into expression (\ref{eqn:deltatau0jagain}) and manipulating the derivatives, we obtain the following expression for the field momentum density:
\begin{equation}
\delta\tau^{0j} = 
 -\frac{1}{2\pi}\epsilon_{jpl}
\left[ S_A^q m_B \left(\frac{1}{r_A}\right)_{,q}\left(\frac{1}{r_B}\right)_{,l}\right]_{,p}+(A \leftrightarrow B)\;.
\label{eqn:tau0jderivs}
\end{equation}
Here $(A \leftrightarrow B)$ means the same expression with labels $A$ and $B$ interchanged.
Notice that this expression for the momentum density is the curl of a vector field; or, equally well,
it can be viewed as the divergence of a tensor field.

The total spin-induced, flowing field momentum  is the integral of expression (\ref{eqn:tau0jderivs}) over the exterior region
$\mathcal E$  (cf.\ Fig.\ \ref{fig:Binary1}).  Using Gauss's law, that volume
integral can be converted into the following integral over the boundary of $\mathcal E$
\begin{eqnarray}
\delta p_{\rm field}^j  &=&  -\frac{1}{2\pi}\epsilon_{jpl}
 S_A^q m_B \int_{\mathcal{\partial E}}\left(\frac{1}{r_A}\right)_{,q}\left(\frac{1}{r_B}\right)_{,l} d\Sigma_p \nonumber \\
&& +(A \leftrightarrow B)\;.
 \label{eqn:pext_bndryE}
\end{eqnarray}
The boundary of $\mathcal E$ has three components: the surface $\mathcal S$ far from the binary
on which we compute the binary's total momentum, and the surfaces $\partial A$ and 
$\partial B$ of bodies $A$ and $B$.  The integral over $\mathcal S$ vanishes because
the integrand is $\propto 1/r^4$ and the surface area is $\propto r^2$ and $\mathcal S$ is
arbitrarily far from the binary, $r\rightarrow \infty$.  When integrating over the bodies' surfaces,
we shall flip the direction of the vectorial surface element so it points out of the bodies (into 
$\mathcal E$), thereby picking up a minus sign and bringing Eq. (\ref{eqn:pext_bndryE}) into
the form
\begin{eqnarray}
\delta p_{\rm field}^j &=& 
 \frac{1}{2\pi}\epsilon_{jpl} S_A^q m_B 
\left[  \int_{\mathcal{\partial A}} 
 \left(\frac{1}{r_A}\right)_{,q}\left(\frac{1}{r_B}\right)_{,l}d\Sigma_p \right. \nonumber \\
 &&  \left. +  \int_{\mathcal{\partial B}} 
\left(\frac{1}{r_A}\right)_{,q}\left(\frac{1}{r_B}\right)_{,l}d\Sigma_p\right] 
+ (A \leftrightarrow B)\,.\nonumber\\
\label{eqn:pjext_bndryAB}
\end{eqnarray}
We presume (as is required by the PN approximation) that the bodies' separation is
large compared to their radii.  Then on $\partial A$, we can write 
$(1/r_A)_{,q} = - n_A^q /r_A^2$ and 
$(1/r_B)_{,l} = n_{AB}^l/r_{AB}^2$, where  $n_A$ is the unit vector pointing away from the center of mass of body $A$, $n_{AB}^l$ is the unit vector pointing from the center of mass of body $B$
toward the center of mass of body $A$, and $r_{AB}$ is the (flat-spacetime) distance between the
two bodies' centers of mass.  The first integral in Eq.\ (\ref{eqn:pjext_bndryAB}) then becomes
$n_{AB}/r_B^2 \int_{\partial A} n_A^q /r_A^2 d\Sigma_p$.  For simplicity we take the surface of integration
to be a sphere immediately above the physical surface of body $A$ and ignore
the tiny contribution from the region between that sphere and the physical surface.
On this sphere, we write $d\Sigma_p = r_A^2 
n_A^p d\Omega_A$, where $d\Omega_A$ is the solid angle element, and we then carry out
the angular integral using the relation
$\int_{\partial A} n_A^q n_A^p d\Omega_A = (4\pi/3) \delta_{qp}$.  Thereby we obtain for the
first integral in (\ref{eqn:pjext_bndryAB}) $ (4\pi/3) \delta_{qp} n_{AB}/r_B^2$ independent
of the radius $r_A$ of the sphere of integration.  (If the body is not spherical, the contribution from
the tiny volume between our spherical integration surface and the physical surface will be 
negligible.)  Evaluating the second integral in Eq. (\ref{eqn:pjext_bndryAB}) in the same way,
and carrying out straightforward manipulations, we obtain for the external field momentum
\begin{equation}
\delta \bm p_{\rm field} = \frac{2}{3 r_{AB}^2} (m_B \bm S_A - m_A \bm S_B)\times \bm n_{AB}\;.
\label{pjext_final}
\end{equation}

For the extreme-kick configuration, which has $m_A = m_B = m$ and $\bm S_B = - \bm S_A$,
this field momentum becomes expression (\ref{eqn:EKdeltapfield}).

\subsection{Centers of Mass and Equation of Motion for the Binary's Compact Bodies}
\label{sec:CMeom}

Restrict attention, temporarily, to a body that is a star rather than a black hole, and temporarily
omit the subscript $K$ that identifies which body.  Then, 
following the standard procedure in special relativity (e.g.\ Box 5.6 of MTW \cite{MTW}), we 
define the star's center-of-mass world line  to be that set of events $x^\mu_{\rm cm}$
satisfying the covariant field-theory-in-flat-spacetime relationship 
\begin{equation}
S^{\alpha\beta}p_\beta = 0\;.
\label{eqn:SPzero}
\end{equation}
Here $p^\beta = \int \tau^{0\beta}d^3x$ is the body's 4-momentum  and
\begin{equation}
S^{\alpha\beta} \equiv \int [(x^\alpha - x^\alpha_{\rm cm})\tau^{\beta 0} 
- (x^\beta - x^\beta_{\rm cm}) \tau^{\alpha 0}] d^3x\;
\label{eqn:SalphabetaDef}
\end{equation}
is the body's tensorial angular momentum.  Here the integrals extend over the
star's interior,  and because the star's momentum is changing, we take the time component
of $x^\mu_{\rm cm}$ to be the same as the time at which the integral is performed,
$x^0_{\rm cm} = x^0$.  (If the momentum were not changing, this restriction would be unnecessary;
cf.\ Box 5.6 of MTW.)  

In a reference frame where the body moves with ordinary velocity $v^j = p^j / p^0$, Eq.\ 
(\ref{eqn:SPzero}) says $S^{i0} = S^{ij} v_j$.  We wish to rewrite this in a more illuminating
form, accurate to first order in the velocity $\bm v$.  At that accuracy, we can evaluate $S^{ij}$ in the body's rest frame, obtaining
$S^{ij} = \epsilon^{ijk} S_k$ where $S_k$ is the body's spin angular momentum
\begin{equation}
S_k = \int \epsilon_{klm} (x^l-x^l_{\rm cm}) \tau^{m0} d^3x\;.
\label{eqn:SpinDef}
\end{equation}
Using definition (\ref{eqn:SalphabetaDef}) of $S^{i0}$ with $x^0_{\rm cm} = x^0$,
our definition (\ref{eqn:SPzero}) of the center of mass then takes the concrete form
\begin{equation}
m \bm x_{\rm cm}  = \int \bm x \tau^{00} d^3 x - \bm v \times \bm S\;.
\label{eqn:xcmp0}
\end{equation}
Here on the left side we have replaced $p^0 = \int \tau^{00} d^3x$ by its value in the body's rest frame, which is the
the mass $m$, since the two differ by amounts quadratic in $\bm v$.   

Notice that, \textit{when computed in the body's rest frame so $\bm v=0$, the center of mass is
$m \bm x_{\rm cm} = \int \bm x \tau^{00} d^3x$, but when computed in any frame moving slowly
with respect to the rest frame, this expression must be corrected by the term
$-\bm v \times \bm S$ that we discussed physically in Sec.\ \ref{sec:EKholemomenta} [Eq.\ 
(\ref{eqn:xcm})].} 
We asserted and used Eq.\ (\ref{eqn:xcmp0}), the ``physical SSC'',  in our analysis of the extreme-kick configuration
[Eq.\ (\ref{eqn:xcmGR})].

In our Harmonic coordinate system and at the 1.5PN order of our analysis, the dominant,
time-time component of the Einstein field equations (\ref{eqn:efe}) reduces to $\eta^{\mu\nu}
{\gg^{00}}_{,\mu\nu} = 16\pi \tau^{00}$.  The type of analysis carried out in Sec.\ 19.1 of MTW \cite{MTW} then
reveals that \textit{in the star's rest frame,
the monopolar part of its $\gg^{00}$ is centered on the location $\bm x_{\rm cm}$;
or, equivalently, when one expands the star's $\gg^{00}$ around $\bm x_{\rm cm}$ in its own
rest frame, there is no dipolar $1/r^2$ term (no mass dipole moment).   This well-known result
(e.g \cite{RMP,ThorneHartle}) can be used as an alternative definition of $\bm x_{\rm cm}$ --- a definition that
works for black holes as well as for stars.}

Using this monopolar-field-centered definition of $\bm x_{\rm cm}$, Thorne and Hartle 
\cite{ThorneHartle} have employed matched asymptotic expansions (valid for black holes)
to derive the equations of motion for a system of compact bodies, e.g.\ a compact binary
[their Eqs.\ (4.10) and (4.11)].  For a compact binary, the spin-induced contributions to these equations of motion at 1.5PN order are [Eq.\ (4.11c)] of Thorne and Hartle]
\begin{eqnarray}
m_A \frac{d\delta \bm v_A}{dt} &=& \frac{m_A}{r_{AB}^3} \left[ 6 \bm n_{AB}(\bm S_B \times \bm n_{AB}
\cdot \bm v_{AB}) + 4 \bm S_B \times \bm v_{AB} \right. \nonumber \\
&&\left. \quad\quad - 6 (\bm S_B \times \bm n_{AB})(\bm v_{AB}\cdot \bm n_{AB})\right] \nonumber \\
 && \hskip-6pt + \frac{m_B}{r_{AB}^3} \left[ 6 \bm n_{AB}(\bm S_A \times \bm n_{AB}
\cdot \bm v_{AB}) + 3 \bm S_A \times \bm v_{AB} \right. \nonumber \\
&&\left. \quad\quad - 3 (\bm S_A \times \bm n_{AB})(\bm v_{AB}\cdot \bm n_{AB})\right] \;.
 \label{eqn:eomTH}
 \end{eqnarray}
Here 
\begin{equation}
\bm v_A \equiv \frac{d\bm x_{{\rm cm}\,A}}{dt}\;, \quad
\delta \bm v_A = \frac{d \delta {\bm x}_{{\rm cm}\, A}}{dt}\;,  \quad
\bm v_{AB} = \bm v_A - \bm v_B
\end{equation}
are the velocity of (the center of mass of) body $A$, the spin-induced perturbation
of that velocity, and  the relative velocity of bodies $A$
and $B$.  The first two lines of Eq.\ (\ref{eqn:eomTH}) are due to frame dragging
by the other body (body $B$); the last two lines are a force due to the coupling of body $A$'s spin to
$B$'s spacetime curvature.

\subsection{The Momenta of the Binary's Bodies}
\label{sec:BodyMomentum}

As in the previous subsection, we initially restrict ourselves to a body that is a star; then we
shall generalize to a black hole; and we initially omit the star's label $A$ or $B$ for ease of notation.

For a star we can derive an expression for the momentum $p^j = \int \tau^{0j} d^3x$ (with the
integral over the star's interior) in terms of the star's
velocity $v^j = dx^j_{\rm cm}/dt$ by differentiating the center-of-mass equation (\ref{eqn:xcmp0}) with
respect to time.  To allow for the possibility that the mass might change with time, we set
$m= \int \tau^{00} d^3 x$ before doing the differentiation; i.e., we differentiate
\begin{equation}
x^j_{\rm cm} \int_A \tau^{00}d^3x = \int_A x^j \tau^{00} d^3 x - (\bm v \times \bm S)^j\;.
\label{eqn:ToDiff}
\end{equation}
Using ${\tau^{00}}_{,0} = - {\tau^{0k}}_{,k}$ and Gauss's theorem, we
bring the left side into the form $v^j \int_A \tau_{00} d^3 x - x^j_{\rm cm} \int_{\partial A} 
(\tau^{0j} - \tau^{00} v^j) d\Sigma_j$.  The last  term arises from the motion
of the surface of the star through space with velocity $\bm v$.  Manipulating the time derivative of the integral on
the right side of Eq.\ (\ref{eqn:ToDiff}) in this same way, we bring it into the form 
$\int_A \tau^{0j} d^3 x - \int_{\partial A} x^j (\tau^{0k} - \tau^{00} v^k ) d\Sigma_k = p^j - \int_{\partial A} x^j (\tau^{0k} - \tau^{00} v^k ) d\Sigma_k $, where $p^j$ is the star's momentum.  Inserting these
expressions for the left side and the right-side integral into Eq.\  (\ref{eqn:ToDiff}), noting that
the star's spin angular momentum evolves (due to precession) far more slowly than its velocity,
denoting the time derivative of its velocity by $d \bm v/dt = \bm a$ (acceleration), solving for
$p^j$ and restoring subscript $A$'s, we obtain
\begin{eqnarray}
p^j_A &=& m_A v^j_A + \int_{\partial A}(x^j - x^j_{{\rm cm}\, A}) (\tau^{0k} - \tau^{00} v^k_A) d\Sigma_k \nonumber \\
&&+ (\bm a_A \times \bm S_A)^j\;.
\label{eqn:pjvint}
\end{eqnarray}

Although we have derived this equation for a star, it must be true also for a black hole.  The reason
is that all the quantities that appear in it are definable without any need for integrating over the
body's interior, and all are expressible in terms of the binary's masses and spins and its bodies'
vectorial separation, in manners that are insensitive to whether the bodies are stars or holes.
To illustrate this statement, in Appendix \ref{App:BHmom} we deduce (\ref{eqn:pjvint}) for a black hole, 
restricting ourselves to spin-induced portion of the momentum that is being exchanged with the
field, $\delta p^j_A$.

It is this $\delta p^j_A$ that interests us.  Because the spin has no influence on $\tau^{00}$ at
the relevant order (which is $\delta\tau^{0k} \sim g H$ and $\delta \tau^{00} \sim g^2$ where $g$
and $H$ are the gravitoelectric and gravitomagnetic fields), 
Eq.\ (\ref{eqn:pjvint}) implies that
\begin{equation} 
\delta p^j_A = m_A \delta v_A^j + \int_{\partial A} (x^j - x^j_{{\rm cm}\, A}) \delta \tau^{0k} d\Sigma_k
+ (\bm a_A \times \bm S_A)^j\;.
\label{deltapA1}
\end{equation}
The acceleration $\bm a_A$ of body $A$ is, at the order needed, just the gravitoelectric field of
body $B$ at the location of $A$, $a_A = -(m_B/r_{AB}^2) \bm n_{AB}$. Performing the surface integral on a sphere just above the body's physical surface we can write $x^j - x^j_{{\rm cm}\,A}
= n_A^j r_A$ and $d\Sigma_k = r_A^2 d\Omega_A$.  Inserting these into Eq.\ (\ref{deltapA1}),
we obtain
\begin{equation}
\delta p^j_A = \underbrace{m_A \delta v_A^j}_{{\mbox{\small kinetic}}\atop {\mbox{\small term}}}
\underbrace{ + \int_{\partial A}  r_A^3 \delta \tau^{0k} n^j_A n^k_A d\Omega_A}_{\mbox{\small surface term}}
\underbrace{+\frac{m_B}{r_{AB}^2} (\bm S_A \times \bm n_{AB})^j }_{\mbox{\small SSC term}}     \;.
\label{deltapA2}
\end{equation} 
Here ``SSC term'' refers to the ``spin supplementary condition'' required to get the correct,
physical center of mass; see text following Eq.\ (\ref{eqn:xcmGR}).   
In the surface term, the field momentum density $\delta \tau^{0k}$ is given by Eq.\ (\ref{eqn:tau0jderivs}).
The second term  $(A \leftrightarrow B)$ is smaller than the first by
$M/r_{AB}$ and thus is negligible.  Inserting the first term into the integral, 
using $(1/r_A)_{,qp} = (3 n_A^q n_A^p - \delta_{qp})/r_A^3$ and $(1/r_B)_{,l} = - n_B^l/r_B^2$, and 
$\int n_A^j n_A^l d\Omega_A = \frac{4\pi}{3} \delta_{jl}$, we bring Eq.\ (\ref{deltapA2}) into the form
\begin{eqnarray}
\bm \delta p_A &=& \underbrace{m_A\delta  \bm v_A}_{{\mbox{\small kinetic}}\atop {\mbox{\small term}}}
\underbrace{ - \frac23 \frac{m_B}{r_{AB}^2} \bm S_A \times \bm n_{AB}}_{\mbox{\small surface term}}
\underbrace{+ \frac{m_B}{r_{AB}^2} \bm S_A \times \bm n_{AB} }_{\mbox{\small SSC term}} 
\nonumber\\
&=& m_A \delta \bm v_A + \frac13 \frac{m_B}{r_{AB}^2} \bm S_A \times \bm n_{AB}\;.
\label{deltapA3}
\end{eqnarray} 

\subsection{Momentum Conservation}
\label{sec:momcons}

The total spin-induced momentum perturbation, $\delta \bm p_{\rm tot} = \delta \bm p_A +
\delta \bm p_B + \delta \bm p_{\rm field}$ [Eqs.\ (\ref{deltapA3}) and (\ref{pjext_final})] is
\begin{equation}
\delta \bm p_{\rm tot} = m_A\delta  \bm v_A + m_B \delta \bm v_B +  \frac{1}{r_{AB}^2}
(m_B  \bm S_A - m_A \bm S_B) \times \bm n_{AB}\;.
\label{eqn:deltaptot}
\end{equation}
Momentum conservation requires that the time derivative of this $\delta \bm p_{\rm tot} $ vanish.
The time derivative of the kinetic terms can be read off the equation of motion (\ref{eqn:eomTH}):
\begin{eqnarray} 
&&m_A \frac{d\delta \bm v_A}{dt} + m_B \frac{d\delta \bm v_B}{dt} 
\label{eqn:KineticDeriv} \\
&&=
- (m_B \bm S_A - m_A \bm S_B) \times [ \bm v_{AB} - 3(\bm n_{AB} \cdot {\bm v}_{AB}) \bm n_{AB} ]
\;.
\nonumber
\end{eqnarray}
By inserting $\bm n_{AB} = (\bm x_{{\rm cm}\,A} - \bm x_{{\rm cm}\,B})/r_{AB}$ into the second term
of Eq.\ (\ref{eqn:deltaptot}) and differentiating with respect to time, we obtain the negative of
expression (\ref{eqn:KineticDeriv}).  Therefore, 
\begin{equation}
d \delta \bm p_{\rm tot} / dt = 0\;;
\label{eqn:deltaMomCons}
\end{equation}
i.e., as the binary's evolution drives spin-induced momentum back and forth between the bodies
and the field, the total momentum remains conserved, as it must.

Interestingly, during the summation of momentum terms, one finds that the surface terms in $\delta \bm p_A + \delta \bm p_B$ have exactly cancelled
the field momentum $\delta \bm p_{\rm field}$, leaving the total momentum as the sum of
the bodies' kinetic term and their SSC term---i.e.\ leaving it equal to the bodies' total
{\it canonical momentum} (see the discussion at the end of Sec.\ \ref{sec:EKMomCons}). 

\section{Conclusion}
In this paper we have explored the flow of momentum between a compact binary's bodies
and their external gravitational field (spacetime curvature), at 1.5PN order, during the binary's
orbital inspiral.  In subsequent papers we shall explore momentum flow in numerical-relativity
simulations of a binary's collision, merger and ringdown.  We expect these studies to give
useful intuitive insights into the internal dynamics of binary black holes and the nonlinear
dynamics of curved spacetime.

\section*{Acknowledgments}   For helpful discussions we thank Geoffrey Lovelace and
Jeff Kaplan.  
This research was supported in part by NSF grants  PHY-0601459
and PHY-0653653, and by the David and Barbara Groce start-up fund at 
Caltech. 

\appendix

\section{The Total PN Momentum Density}
\label{sec:MomDenFull}

In Section \ref{sec:FieldMom}, Eqs.\ (\ref{eqn:deltatau0j}) and (\ref{eqn:EKfieldmom}) show the portion of the 
field momentum that generates bobbing.
There are, however, additional pieces of field momentum at the same PN order that do not contribute to the bobbing,
and these expressions become important for comparisons of the post-Newtonian analysis with numerical-relativity
results (to be presented in future papers).
There are three extra sources of terms.
First, the gravitomagnetic field has a part  $\bm H^{\rm velo}$ that depends upon the body's velocity.
Using Eqs.\ (2.5) and (6.1) of \cite{KNT}, one can see that
\begin{equation}
	\bm H^{\rm velo} = \frac{4 m_A ({\bf n}_A \times {\bf v}_A)}{r_A^2} + (A \leftrightarrow B)
\end{equation}
Second, there are terms from the coupling of the gravitomagnetic field of body $A$
with its own gravitoelectric field  $\tau^{0j} \bm e_j = -(\bm g_A \times \bm H_A)/(4\pi)$,
and similarly for body $B$, for both the spin and velocity pieces of $\bm H$.
Finally, there is a part due to $(3\dot U_N \bm g)/(4\pi)$, where $U_N$ is the
Newtonian potential and the dot denotes differentiation with respect to time (see Eq. (4.1)
of \cite{KNT}).
When one accounts for these additional expressions, the full field momentum density is written most concisely as
\begin{subequations}
\begin{equation}
\tau^{0j} = \tau^{0j}_{\rm spin} + \tau^{0j}_{\rm velo},
\label{eqn:tautot}
\end{equation}
where $\tau^{0j}_{\rm spin}$ and $\tau^{0j}_{\rm velo}$ are the terms that depend upon the spins and the velocities, 
respectively.
These terms are given by
\begin{eqnarray}
	\tau^{0j}_{\rm spin}{\bm e}_j & = & \frac{m_B}{2\pi r_A^3 r_B^2} \left[ 3({\bf S}_A \cdot {\bf n}_A)({\bf n}_A 
	\times {\bf n}_B) - ({\bf S}_A \times {\bf n}_B)\right]\nonumber\\
	&& - \frac 1{2\pi}\frac{m_A}{r_A^5}({\bf S}_A\times{\bf n}_A) + (A\leftrightarrow B),
	\label{eqn:tauspin}
\end{eqnarray}
and
\begin{eqnarray}
	&&\tau^{j0}_{\rm velo}{\bf e}_j = \frac{m_A}{4\pi r_A^2}\left\{\frac{m_B[4({\bf n}_B\cdot
	{\bf v}_A){\bf n}_A - 4({\bf n}_A\cdot{\bf n}_B){\bf v}_A]}{r_B^2} \right.\nonumber\\
	&& \quad\quad\left.- \frac{3m_B({\bf n}_A\cdot{\bf v}_A){\bf n}_B}{r_B^2} +\frac{m_A[({\bf n}_A \cdot {\bf v}_A){\bf n}_A 
	- 4{\bf v}_A]}{r_A^2} \right\}\nonumber\\
	&&\quad\quad+(A\leftrightarrow B).
\end{eqnarray}
\end{subequations}
In the body of this paper we have confined attention to the first line of Eq.\ (\ref{eqn:tauspin})
and its $(A \leftrightarrow B)$, as that is the part of the field momentum that gets exchanged with the bodies during bobbing.

\section{Momentum of a Black Hole}
\label{App:BHmom}

In the text we derived expression (\ref{eqn:pjvint}) for the momentum of a body in a binary, assuming the
body is a star so we could do volume integrals, and we then asserted that this expression is
also valid for black holes.  
The spin-induced portion of this expression that gets exchanged with the field as the body moves is given by Eq.\ (\ref{deltapA1}), which reduces to (\ref{deltapA3}).  
In this appendix we shall sketch a derivation of Eq.\ (\ref{deltapA3}) directly from the surface-integral definition
(\ref{eqn:pAsurf}) of a black hole's momentum,
\begin{equation}
	\delta p^j_A = \frac 1{16\pi} \int_{\partial A} \delta {H^{j\alpha0k}}_{,\alpha} d\Sigma_k\;.
\label{eqn:deltapAsurfint}
\end{equation}
To evaluate this surface integral up to  desired 1.5PN-order accuracy turns out to require some 2.5PN   fields. Qualitatively, this can be anticipated because the superpotential we use in the surface integral is {\it sourced} by the spin-orbit piece of field momentum, and therefore necessarily a non-leading PN term.  One can see this more clearly by expanding $\delta {H^{j\alpha0k}}_{,\alpha}$ in terms of the metric density 
and using the symmetries of the superpotential $H$ (which are the same as the Riemann tensor).
In general, the momentum is given by
\begin{equation}
	\delta p^j_A = -\frac 1{16\pi} \int_{\partial A} (\gg^{jk}\gg^{\alpha 0} - \gg^{j\alpha}\gg^{0k})_{,\alpha} 
	d\Sigma_k.
\end{equation}
In Harmonic gauge, however, $\gg^{\alpha\beta}{}_{,\beta}=0$, and the spatial metric is flat until 2PN order while
the time-space components are of 1.5PN order.
As a result, the terms at lowest and next-to-lowest PN order are contained within two terms,
\begin{equation}
	\delta p^j_A = \frac 1{16\pi} \int_{\partial A} (\gg^{j0}{}_{,k}+\gg^{jk}{}_{,0}) d\Sigma_k.
	\label{eqn:momharmgauge}
\end{equation}
In this  expression, the momentum arises from linear terms involving the metric density, instead of quadratic ones.  As a result, one must keep pieces of the metric perturbation that are of higher PN accuracy.  (Note: if we were to evaluate the time derivative of $\delta \bm p_A$ using the surface
integral (\ref{eqn:dpAdtsurf}), we would not face such a delicacy; the integrand there is quadratic and
requires only 1.5PN fields for its evaluation.)

To find the momentum in terms of the standard post-Newtonian parameters, we resort to a standard way  that the metric perturbations are written in recent post-Newtonian literature,  e.g., by Blanchet, Faye, and Ponsot \cite{BFP}:
\begin{subequations}
\begin{eqnarray}
	g_{00} & = & -1 + 2V - 2V^2 + 8 \hat X\\
	g_{i0} & = & -4V_i -8\hat R_i\\
	g_{ij} & = & \delta_{ij}(1 + 2V + 2V^2) + 4\hat W_{ij}\\
	\sqrt{-g} & = & 1 + 2V + 4V^2 + 2\hat W_{kk}.
\end{eqnarray}
\end{subequations}
For spinning systems, we adopt the notation of Tagoshi, Ohashi, and
Owen \cite{TOO} where $\mathcal{O}(m,n)$ means to order $c^m$ for non-spinning 
terms and $\chi c^n$ for terms involving a single spin $\chi$. (Here $\chi = |\bm S|/m^2$ is the
body's dimensionless spin.)  In this notation, 
terms we are interested in are of the order 
$\mathcal{O}(3,6)$, while the above post-Newtonian potentials have been obtained
up to the 
following orders~\cite{TOO,Faye}:  
\begin{eqnarray}
V=\mathcal{O}(2,5), & V_{j}=\mathcal{O}(3,4), \nonumber \\
\hat{W}_{jk}=\mathcal{O}(4,5), & \hat{R}_{j}=\mathcal{O}(5,6).
\end{eqnarray}
In terms of these post-Newtonian potentials,  $V$, $V_i$, $\hat R_i$, $\hat X$ and $\hat W_{ij}$, the perturbed metric density is
\begin{subequations}
\begin{eqnarray}
	\gg^{00} & = & -1 - 4V -2\left( \hat W_{kk} + 4V^2 \right) + \mathcal{O}(6,7)\quad \\
	\gg^{0i} & = & -4V_i - 8\left( \hat R_i + VV_i \right) + \mathcal{O}(6,7) \quad \\
	\gg^{ij} & = & \delta_{ij} - 4\left( \hat W_{ij} - \frac{1}{2}\delta_{ij}\hat W_{kk}\right)+ \mathcal{O}(6,7).\quad
\end{eqnarray}
\end{subequations}
As a consequence, Eq.\ (\ref{eqn:momharmgauge}) is given by 
\begin{eqnarray}
	\delta p_A^j & = & \frac 1{16\pi} \int_{\partial A} \bigg\{ \left[-4V_{j} - 8
	\left( \hat{R}_{j(S)} + V_{(m)}V_{j(S)}\right)\right]_{,k}  \nonumber \\
	& & - 4\left[\hat W_{jk(S)} - \frac{1}{2}\delta_{jk} \hat{W}_{ii(S)}\right]_{,0} \bigg\} d\Sigma_k\,,
\label{eqn:SuperPotentialTagoshi}
\end{eqnarray}
where a subscript $(S)$ means keep only the parts of those potentials proportional to the spins of the bodies,
and a subscript $(m)$ involves pieces of the potential without spins (proportional to the masses of the bodies).
Terms without a subscript have both pieces.

Tagoshi, Ohashi and Owen express the potentials $V_{(m)}$, $V_{j}$, $\hat R_{j(S)}$ and $\hat W_{jk(S)}$ in terms of
the bodies' masses, vectorial velocities, vectorial spins, and vectorial separations, and distance to the field-point 
location [their Eqs.\  (A1a), (A1d), (A1f), and (A1g)].
While the full equations are quite lengthy, the portions that generate momentum flow -- those involving
the coupling of the mass of one body to the spin of the other -- are somewhat simpler.
For convenience, we give these portions of the equations below, rewritten in our notation, with the typos noted 
by G.\ Faye et al \cite{Faye} corrected.
\begin{subequations}
\begin{eqnarray}
	V_{(m)} & = & \frac{m_A}{r_A} + (A \leftrightarrow B)\\ 
	\nonumber
	V_{j} & = & \frac{m_A v_A^j}{r_A}+\epsilon_{jkl}S_A^k \left\{n_A^l\left[-\frac{3m_B}{2r_A^2r_{AB}}\right.\right.\\
	&&\left.\left. - \frac{m_B({\bf n_A}\cdot{\bf n_{AB}})}{4r_A r_{AB}^2} \right]
	+ n_{AB}^l\frac{3m_B}{4r_Ar_{AB}} \right\}\\
	\nonumber
	\hat W_{jk(S)} & = & \left[ \frac 12 \left( \epsilon_{ijl} S_A^l v_A^k + \epsilon_{ikl} S_A^l v_A^j \right) -
	\delta_{jk} \epsilon_{ilm} v_A^l S_A^m \right]\\
	&& \times \frac{n_A^i}{r_A^2} + (A \leftrightarrow B)\\
	\nonumber
	\hat R_{j(s)} & = & \epsilon_{jkl} S_A^k \left[ n_A^l \left( -\frac{m_B}{2r_A^2r_{AB}} + 
	\frac{m_B}{r_{AB}s^2} \right) \right.\\
	\nonumber
	&& n_{AB}^l \left( -\frac{m_B}{2r_Ar_{AB}^2} + \frac{m_B}{2r_{AB}^2r_B} + \frac{m_B}{r_A s^2} \right)\\
	\nonumber
	&& \left. n_B^l \left( \frac{m_B}{r_As^2} + \frac{m_B}{r_{AB}s^2} \right)\right]\\
	\nonumber
	&& n_A^j \epsilon_{ikl} S_A^l\left[n_A^i(n_{AB}^k + n_B^k)\left(\frac{m_B}{r_As^2} + \frac{2m_B}{s^3}\right)\right.\\
	\nonumber
	&& \left. -2n_{AB}^i n_B^k \frac{m_B}{s^3} \right] + n_{AB}^j \epsilon_{ikl}S_A^l\left[-2n_A^i n_b^k \frac{m_B}{s^3}
	\right.\\
	\nonumber
	&& \left. (n_A^i+n_B^i)n_{AB}^k\left(-\frac{m_B}{r_{AB}s^2}-\frac{2m_B}{s^3}\right)\right]\\
	&&+ (A \leftrightarrow B),
\end{eqnarray}
\end{subequations}
Here, as before, $m_A$, $\bm v_A$, and $\bm S_A$ are the mass, velocity, and spin angular-momentum of object $A$; 
$r_A$ is the separation of body $A$ from a point in space and $r_{AB}$ is the separation of the two objects;
and $\bm n_A$ and $\bm n_{AB}$ are unit vectors pointing along $r_A$ and $r_{AB}$, respectively.
A new quantity, $s=r_A+r_B+r_{AB}$, has been introduced, in addition.
Inserting these expressions into Eq.\ (\ref{eqn:SuperPotentialTagoshi}) gives us $\delta H^{j\alpha 0 k}{}_{,\alpha}$,
and the momentum of body $A$ is then found by performing a surface integral over $A$'s surface.
The surface integrals are computed under the same assumptions as in Section \ref{sec:ExtFieldMom}; namely the separation
of the bodies is much larger than their radii, and each surface of integration is a sphere immediately above a body's surface. 
When they are computed, one finds the same result as Eq.\ (\ref{deltapA3}),
\begin{equation}
	\delta \bm p_A =  m_A \delta \bm v_A + \frac13 \frac{m_B}{r_{AB}^2} \bm S_A \times \bm n_{AB}.
	\label{eqn:surfpA}
\end{equation}
One can find the momentum for body $B$ by exchanging  $A$ and $B$.

As a consistency check, we can evaluate the system's total momentum by doing a surface integral 
at infinity: 
\begin{equation}
	\delta {p}^j_{\rm tot} = \frac{1}{16\pi}\oint_{\mathcal S} \delta {H^{j\alpha 0 k}}_{,\alpha} d\Sigma_k
\end{equation}
The quantity $\delta {H^{j\alpha 0 k}}_{,\alpha}$ is exactly the same as above, from which one can find
\begin{equation}
	\delta {p}^{j}_{\rm tot} = m_A \delta v_A^j + \frac{m_B\left(\mathbf{S}_A \times \mathbf{n}_{AB}\right)^j}{r_{AB}^2} 
	+ (A \leftrightarrow B)
	\label{eqn:surfptot}
\end{equation}
This, combined with the fact that 
\begin{equation}
	\delta \bm p_{\rm tot} = \delta \bm p_A +\delta  \bm p_B +\delta  \bm p_{\rm field}.
	\label{eqn:surfmomcons}
\end{equation}
as well as Eq.~\eqref{eqn:surfpA}, gives
\begin{equation}
\delta  {p}^{j}_{\rm field} = \frac{2 m_B\left({\mathbf{S}_A \times \mathbf{n}_{AB}}\right)^j}{3 r_{AB}^2} + (A \leftrightarrow B),
\end{equation}
as found in Section \ref{sec:ExtFieldMom}.


\begin{thebibliography}{10}

\bibitem{Pretorius1} F.\ Pretorius, Phys.\ Rev.\ Lett.\ {\bf 95}, 121101 (2005).
%
\bibitem{DynamicalHorizons} A.\ Ashtekar and B.\ Krishnan, 
Living Rev.\ Relativity {\bf 7}, 10.  URL (cited on 5 February 2009):
http://www.livingreviews.org/lrr-2004-10  (2008).
%
\bibitem{Quasilocal} L.B.~Szabados, Living Rev.\ Relativity {\bf 7}, 4. (2004).  URL (cited on February 21, 2009):  http://www.livingreviews.org/lrr-2004-4.
%
\bibitem{CNT}  C.-M.\ Chen, J.M.\ Nester, and R.S. Tung, Phys.\ Rev.\ D {\bf 72}, 104020 (2005).
%
\bibitem{CLZM1} M.\  Campanelli, C.\ O.\ Lousto, Y.\ Zlochower, and D.\ Merritt, Astrophys.\ J.\
Lett.\ {\bf 659}, L5 (2007).
%
\bibitem{CLZM2} M.\  Campanelli, C.\ O.\ Lousto, Y.\ Zlochower, and D.\ Merritt, Phys.\ 
Rev.\ Lett.\ {\bf 98}, 231102 (2007).
%
\bibitem{HGSBH} J.\ A.\  Gonzalez, U.\ Sperhake, B.\ Bruegmann, M.\ Hannam and S.\ Husa,
Phys.\ Rev.\ Lett.\ {\bf 98}, 091101 (2007).
%
\bibitem{Healy} J.~Healy et al., Phys.\ Rev.\ Lett.\  {\bf 102}, 041101 (2009).
\bibitem{LL62} L.\ D.\ Landau and E.M.\ Lifshitz, {\it Classical Theory of Fields} (Addison Wesley,
Redding Mass., 1962), Sec.\ 100.
\bibitem{MTW} C.\ W.\ Misner, K.\ S.\ Thorne and J.\ A.\ Wheeler, {\it Gravitation} (Freeman,
San Francisco, 1973).
\bibitem{Babak} S.\ V.\ Babak and L.\ P.\ Grishchuk, Phys.\ Rev.\ D {\bf 61},
024038 (2000).
\bibitem{Pretorius2} F.\ Pretorius, in {\em Relativistic Objects in Compact Binaries: From Birth to Coalescence,} edited by Colpi et al., Springer Verlag (see Sec.\ IV.C).  \url{http://xxx.lanl.gov/abs/0710.1338}
\bibitem{Vortices} A.\ H.\ Shapiro, {\it Vorticity}, a film in a series produced by the National
Committee on Fluid Mechanics (1961); available for streaming at \url{http://web.mit.edu/hml/ncfmf.html};
segment 4 minutes 26 seconds into Part II.
\bibitem{ThorneHartle} K.\ S.\ Thorne and J.\ B.\ Hartle, Phys.\ Rev.\ D {\bf 31}, 1815 (1985).
\bibitem{KNT} J.\ D.\ Kaplan, D.\ A.\ Nichols and K.\ S.\ Thorne, Phys.\ Rev.\ D to be submitted, \url{http://xxx.lanl.gov/abs/0808.2510}
\bibitem{PatiWill} M.\ E.\ Pati and C.\ M.\ Will, Phys.\ Rev.\ D {\bf 62}, 124015 (2000).
\bibitem{BT} R.\ D.\ Blandford and K.\ S.\ Thorne, {\it Applications of Classical Physics}, 
textbook nearing completion and available at \url{http://www.pma.caltech.edu/Courses/ph136/yr2008} .
\bibitem{SSC} see e.g., Appendix A of L.E.~Kidder, Phys.~Rev.~D {\bf 52}  821 (1995). 
\bibitem{DJS} T.\ Damour, P.\ Jaronowski and G.\ Sch\"afer, Phys.\ Rev.\ D {\bf 77}, 064032 (2007).
\bibitem{RMP} K.\ S.\ Thorne, Rev.\ Mod.\ Phys {\bf 52}, 299 (1980).
\bibitem{BFP} L.\ Blanchet, G.\ Faye, and B.\ Ponsot, Phys.\ Rev.\ D {\bf 58}, 124002 (1998).
\bibitem{TOO} H.\ Tagoshi, A.\ Ohashi, and B.\ J.\ Owen, Phys.\ Rev.\ D {\bf 63}, 044006 (2001).
\bibitem{Faye} G.\ Faye, L.\ Blanchet, and A.\ Buonanno, Phys.\ Rev.\ D {\bf 74}, 104033 (2006).









\end{thebibliography}
\end{document}